\documentclass[aps,prd,reprint,preprintnumbers,floatfix,superscriptaddress,nofootinbib,longbibliography]{revtex4-2}

\usepackage{dcolumn}

\usepackage{hyperref}
\hypersetup{colorlinks=true}

\usepackage{float}

\usepackage{tabularx}
\usepackage{xcolor}
\usepackage{bookmark}
\usepackage{amsmath}
\definecolor{color_01}{rgb}{0,0,0.75}
\definecolor{color_02}{rgb}{0.75,0,0}
\definecolor{color_03}{rgb}{0.5,0,0.75}
\hypersetup{linktoc=page,linkcolor=color_01,citecolor=color_02,urlcolor=color_03}

\usepackage{graphics}
\usepackage{booktabs}
\usepackage[utf8]{inputenc}
\usepackage[T1]{fontenc}
\usepackage[english]{babel}
\usepackage{mathtools}
\usepackage{amssymb}
\usepackage{bm}
\usepackage{orcidlink}
\usepackage{graphicx}

\begin{document}

\title{Virial theorem for a cloud of stars obtained from Jeans equations \\ with  the second correlation moments}

\author{Stupka A.A.\orcidlink{0000-0003-1404-3899}}
\email{stupkaanton@gmail.com}
 \affiliation{Oles Honchar Dnipro National University, Haharin 72 ave., Dnipro 49010, Ukraine}
\author{Kopteva E.M.\orcidlink{0000-0001-8364-0481}}
 \affiliation{Institute of Physics, Research Centre of Theoretical Physics and Astrophysics, \\ Silesian University in Opava, 746 01 Opava, Czech Republic}%
\affiliation{University of Illinois at Urbana Champaign, Department of Physics, \\ 1110 West Green St., Urbana, IL 61801, USA}%
\author{Saliuk M.A.\orcidlink{0000-0002-6143-350X}}
\affiliation{Oles Honchar Dnipro National University, Haharin 72 ave., Dnipro 49010, Ukraine}%
\author{Bormotova I.M.\orcidlink{0000-0002-7582-7817}}
\affiliation{Institute of Physics, Research Centre of Theoretical Physics and Astrophysics, \\ Silesian University in Opava, 746 01 Opava, Czech Republic}%
\affiliation{Institute of Experimental and Applied Physics, Czech Technical University in Prague,\\ 128 00 Prague, Czech Republic}%

\date{\today}

\begin{abstract}
A hydrodynamic model for small acoustic oscillations in a cloud of stars is built, taking into account the self-consistent gravitational field in equilibrium with a non-zero second correlation moment. It is assumed that the momentum flux density tensor should include the analog of the anisotropic pressure tensor and the second correlation moment of both longitudinal and transverse gravitational field strength. The non-relativistic temporal equation for the second correlation moment of the gravitational field strength is derived from the Einstein equations using the first-order post-Newtonian approximation. One longitudinal and two transverse branches of acoustic oscillations are found in a homogeneous and isotropic star cloud. The requirement for the velocity of transverse oscillations to be zero provides the boundary condition for the stability of the cloud. The critical radius of the spherical cloud of stars is obtained, which is precisely consistent with the virial theorem.
\end{abstract}

\maketitle

\section{Introduction}
\label{intro}
Observed stellar clusters demonstrate a broad spectrum of masses, sizes, and other properties. Current compilations of observational data for clusters show correlations between their mass, size, age, maximum stellar mass, etc. Most theoretical attempts to explain these correlations focus on the stage of star formation in such clusters \cite{Pfalzner,Dib,Parmentier}. However, the statistical effects of the system's evolution at later stages, when stars are formed, can also be of interest, especially for larger systems like super star clusters, globular clusters, or elliptical galaxies.

In this paper, we aim to generalize an approach for studying such statistical effects, considering the transverse part of gravitational field strength and an analog of anisot\-ropic pressure for the gravitating gas of stars. To achieve this, we consider an idealized system — a very large, homogeneous, and isotropic cloud of stars in a stable state. Switching to the non-relativistic limit and using the first-order post-Newtonian (1PN) approximation, we assume that the gravitational field is self-consistent, so that, in equilibrium, it has a zero first correlation moment and a non-zero second correlation moment. By incorporating analogs of anisotropic pressure and turbulent-like motion into our analysis, we investigate low-frequency acoustic oscillations with small amplitudes in the cloud of stars. Additionally, we apply the Jeans theory to analyze the stability of the system.

In the literature, it is common to utilize hydrodynamic formalism to describe large stellar systems (see, e.g., \cite{Lars,Alex}). Recently, the Jeans instability has been analyzed within the framework of the Boltzmann and Poisson equations derived from the 1PN theory \cite{k21,n-k-r-a}. In these papers, small relativistic corrections were found to the Jeans wavelength \cite{j15}
\begin{equation}
k_J=\frac{\sqrt{4 \pi G \rho_0}}{\sigma},
\end{equation}
where $G$ is the gravitational constant, $\rho_0$ is the equilibrium value of the density of the gas of stars, and $\sigma$ is the velocity dispersion.

The account of turbulence for the generalization of the Jeans wavelength estimation was first proposed by Chandrasekhar in \cite{cha}. In the non-relativistic case, when closing the system of Jeans equations, Chandrasekhar utilized the Newtonian approximation of the theory of gravity. According to his estimation, it follows that gravitational instability of small perturbations arises for wavelengths exceeding $\lambda_C$:

\begin{equation}
 \label{cha} \lambda_C=\sqrt{\frac{\pi(v_s^2+
\frac{1}{3}\left\langle {v^2 } \right\rangle_0)}{G\cdot \rho_0}},
\end{equation}
where $v_s = \sqrt{\left( {\partial P/\partial \rho } \right)_s}$ is the speed of sound, $P$ and $\rho$ are the pressure and the density in the system, respectively, $\sqrt{\left\langle {v^2 } \right\rangle_0} $ is the mean-square velocity of the turbulence.

However, we will further demonstrate that our approach provides more appropriate corrections to the Jeans theory in the post-Newtonian approximation. This approach has been effectively applied in describing a giant molecular cloud in \cite{s-k}. The paper showed that, in the isotropic case, the non-relativistic limit of the Einstein equations for transverse field modes yields a larger growth increment compared to the Newtonian theory. The present work extends the considerations carried out in \cite{s-k} to the case of a gas of stars.

In this study, we derive an equation for the second correlation moment of the gravitational field strength in the non-relativistic approximation of Einstein's theory, considering both the longitudinal and transverse parts of the gravitational field. We establish the stability condition for a gravitating homogeneous isotropic stellar system and demonstrate that precisely the transverse perturbations determine the Jeans radius. This stability condition for the gravitating system explicitly satisfies the virial theorem, in contrast to theories that do not consider transverse perturbations, such as \cite{k21} and \cite{cha}.

The further paper is organized as follows. Section~\ref{sec:1} presents the derivation of the wave equation describing the acoustic oscillations in the cloud of stars. In Section~\ref{sec:3}, the acoustic oscillations in the cloud are examined using the non-relativistic 1PN approximation of the Einstein equations, taking into account the self-consistent gravitational field. Section~\ref{sec:4} focuses on the analysis of the stability of the star cloud. The obtained results are then applied to observational data for star clusters. Finally, Section~\ref{conc} concludes the paper.

\section{Jeans Equations with the second correlation moments} \label{sec:1}

To analyse the stability of the homogeneous and isotropic cloud of stars, we use one-particle distribution function $f(\textbf{x},\textbf{v},t)$, which satisfies the collisionless Boltzmann equation (see e.g. \cite{king}, Eq. (1.2))

\begin{equation}\label{f}
\frac{\partial f}{\partial t}+{v_i}\partial_i f+\frac{\partial f}{\partial { v_i}}{F_i}=0.
\end{equation}

Here and throughout the paper, bold letters denote three-vectors in ordinary three-dimensional Euclidean space. Latin indices take on values 1, 2, 3 and represent spatial coordinates. Repeated indices are summed. $\partial_k$ represents the partial derivative with respect to the $k$-th coordinate: $\partial_k=\partial /\partial x_k$. ${F_i}$ denotes the components of of the gravitational interaction force. It is convenient to express the force in terms of the derivative from the momentum flux density of the gravitational field, $\pi^g_{ik}$:

\begin{equation}\label{F}
F_i=\partial_k \pi^g_{ik}.
\end{equation}

The momentum flux density of the gravitational field is defined as follows

\begin{equation}\label{mt1}
\pi^g_{ik} = \frac{\frac{1}{2}\delta _{ik}\left( {g_l g_l } \right)-g_k
g_i}{4 \pi G},
\end{equation}
where $g_i$ represents the components of the gravitational field strength ${\bf g}$.
In the expression (\ref{mt1}), the 1PN approximation for the Landau-Lifshitz pseudotensor is used so that the gravitomagnetic field and time derivatives of the Newtonian potential are neglected (see \cite{k-n-t}, Eq. (4.1b)). As it will be shown below, the gravitational field strength ${\bf g}$ is not a potential vector.

The self-consistent gravitational field in the cloud is a random quantity. After averaging over a physically small spatial value, the first correlation moment of the gravitational field strength equals zero, unlike the second moment, which is non-zero.

Using the definition (\ref{F}), we rewrite (\ref{f}) in the form
\begin{equation}\label{f1}
\frac{\partial f}{\partial t}+{v_i}\partial_i f+\frac{\partial f}{\partial { v_i}}\partial_k \pi^g_{ik}=0.
\end{equation}

The Boltzmann equation (\ref{f1}) provides equations for the velocity correlation moments. Introducing the definition for the mass density
\begin{equation}
\rho=m\int f d^3v,
\end{equation}
and the mass velocity components
\begin{equation}
\rho v_i=m\int v_i f d^3v,
\end{equation}
where $m$ is the average star mass, we then take the first (continuity equation) and the second moments of (\ref{f1}). Thus, we derive the following Jeans equations \cite{j15,b-s}:

\begin{equation}\label{0}
\partial_t \rho + \partial_i(\rho v_i)=0,
\end{equation}

\begin{equation}\label{6g}
\partial_t \left({\rho v_i}\right) + \partial_k \pi_{ik}=0,
\end{equation}

\begin{equation}\label{mt2}
\pi_{ik} = \rho\langle{v_i v_k}\rangle + \frac{\langle g_k
g_i\rangle - \frac{1}{2}\delta_{ik}\langle{g_l g_l}\rangle}{4\pi G},
\end{equation}
where $\pi_{ik}$ denotes the total momentum flux density of the system. Equation (\ref{6g}) corresponds to the equation (4--27) in \cite{b-s}, but taking into account the transverse gravitational field.

In the equation (\ref{6g}), the force of gravitational interaction is entirely defined by the second correlation moment of the gravitational field strength $\left\langle{g_i g_k}\right\rangle$.

Here, $\langle{v_i v_k}\rangle=\overline{v_i v_k}+v_i v_k$ represents the velocity second correlation moment, and $\overline{v_i v_k}=\sigma_{ik}$ denotes the velocity central second correlation moment, known as the velocity dispersion. Similarly, the second correlation moment of the gravitational field strength is $\langle{g_i g_k}\rangle=\overline{g_i g_k}+g_i g_k$, with $\overline{g_i g_k}$ representing the central second correlation moment of the gravitational field strength. Drawing an analogy with the theory of turbulence, the term $-\rho\overline{v_i v_k}$ can be viewed as a stress tensor that describes the anisotropic pressure and corresponds to the Reynolds stress tensor in the theory of turbulence \cite{sh-d}.

It is a common approach to neglect pressure anisotropy in (\ref{6g}) and instead consider ordinary pressure that satisfies the state equation. However, one should not assume the flow in the cloud of stars is laminar. Without viscosity to induce laminarity, nonstationary flows in such a system become vortical. Additionally, when considering both the potential and transverse gravitational fields, the corresponding order accuracy necessitates the inclusion of transverse anisotropic effects in the stress tensor of the medium.

By multiplying (\ref{f1}) by $v_i v_k$, we obtain a temporal equation for the velocity second correlation moment. In equilibrium, the homogeneous and isotropic system is characterized by the following parameters: $\rho_0=\mathrm{const}, \left\langle{v_i v_k }\right\rangle_0=\mathrm{const},$ $\left\langle{g_i g_k }\right\rangle_0=\mathrm{const}$, and $v_0=0$. Our interest lies in small deviations from these equilibrium values, so we derive from (\ref{f1})

\begin{equation}\label{63}
\partial_t \rho\left\langle { v_i v_k} \right\rangle = -\partial_l\rho\langle v_i v_k v_l\rangle.
\end{equation}
Here, the third correlation moment arises, which, due to isotropy, can be expressed in the form 
\begin{equation}
\langle v_i v_k v_l\rangle=v_i\langle v_k v_l\rangle_0+v_k\langle v_i v_l\rangle_0+v_l\langle v_i v_k \rangle_0.
\end{equation}
In the case of isotropy, the equilibrium second correlation moment of the velocity has a tensor structure:

\begin{equation}\label{6v}
\left\langle{v_l v_m}\right\rangle_0 =
\frac{1}{3}\left\langle{v^2}\right\rangle_0\delta_{lm} = \mathrm{const}.
\end{equation}

Thus, (\ref{63}) with account of (\ref{0}) gives the following temporal equation
\begin{equation}\label{rein}
\partial_t\left\langle{v_i v_k}\right\rangle = - \frac{1}{3} \partial_l(v_i\delta_{kl}+v_k\delta_{il})\left\langle{v^2}\right\rangle_0.
\end{equation}

The equation (\ref{rein}) is analogous to the Reynolds stress transport equation in turbulence theory. It describes the transfer of kinetic energy from the mean flow to the small-scale fluctuating motions.

\section{Non-relativistic limit for Einstein Equations}
\label{sec:3}
Treating the gravitational field as a random variable, we will further derive the temporal equation for the quantity $\left\langle{g_i g_k}\right\rangle$ from the Einstein equations.

Let us formulate the conditions for transition to the weak field approximation \cite{ein,m-t-u}. The metric, composed of the Minkowski metric $\eta_{\alpha\beta}$ with small deviations ${h_{\alpha \beta}}\ll 1$, is given by
\begin{equation}\label{mink}
g_{\alpha\beta} = \eta_{\alpha\beta} + h_{\alpha\beta},
\end{equation}
where Greek indices take on values 0, 1, 2, 3 and represent the components of 4-vectors. The zero component of the partial derivative is $\partial_{0}=\frac{1}{c}\partial_{t}$.

In the non-relativistic case, all velocities are much less than the speed of light $c$, and ${\partial_i}{h_{\alpha\beta}}\gg{\partial_0}{h_{\alpha\beta}}$.

In the 1PN approximation (see e.g., \cite{k-n-t}, Eq. (2.5a)) the "electric-like" part of gravitational field is described by the strength
\begin{equation}
\frac{g_i}{c^2}=\frac{1}{2}\partial_i g_{00}-\partial_0 g_{0i} \label{g},
\end{equation}
and "magnetic-like" part (\cite{k-n-t}, (2.5b)) is given by
\begin{equation}
\frac{H_i}{c^2}=\varepsilon_{ijk}\partial_j g_{0k} \label{H}.
\end{equation}

To express the Einstein equations in the 1PN approximation, it is convenient to adopt the metric component definition provided in \cite{DSX} (Eqs. (3.3)). Subsequently, following \cite{k-n-t} (Eq. (2.6d)), it becomes possible to formulate the temporal equation for the gravitational field strength.

To write the Einstein equations in the 1PN approximation, it is
convenient to use the metric components definition provided in
\cite{DSX} (Eqs. (3.3)). Subsequently, following \cite{k-n-t} (Eq. (2.6d)), it is possible to derive the temporal equation for the gravitational field strength in the following form
\begin{equation}\label{new3}
{\partial_t}g_i -\frac{c}{4} [\nabla \times {\bf H}]_i = 4\pi G{{{\rho v_i}}},
\end{equation}
where $\nabla$ is a vector operator with components $\partial_i$.

Multiplying (\ref{new3}) by $g_k$ at the same point of the space-time and making the symmetrization by tensor indexes, we find
\begin{eqnarray}\label{4b}
\nonumber\partial_t(g_i g_k) = &&\frac{c}{4}[\nabla \times{\bf H}]_i g_k
\nonumber\\&&+\frac{c}{4}[\nabla \times {\bf H}]_k g_i + 4\pi G \rho
\left({v_i g_k + v_k g_i }\right).
\end{eqnarray}

We will now perform the statistical averaging of equation (\ref{4b}). To accomplish this, we need to determine the contribution of the gravitomagnetic field resulting from the motion of particles with velocity ${\bf v}$ in the Newtonian gravitational field $\mathbf{g_\mathrm{N}}$, whose components are given by 
\begin{equation}
 g_{\mathrm{N}i}=\frac{1}{2}\partial_i g_{00}c^2.   
\end{equation}
Since the values $(g_{00}, g_{0i})$ form a 4-vector, we transition to the intrinsic frame of reference of the medium, where the reference frame moves with velocity ${\bf v}$ at a given point.

In the non-relativistic approximation, the Lorentz transformations give the relations 
\begin{equation}
 g'_{0i}=g_{0i}-u_i g_{00}, \quad {u_i} = -v_i/c.  
\end{equation} 
The derivatives with respect to the spatial coordinates do not change under these transformations. Thus, we can find the expression for the gravitomagnetic field of the system at rest: 
\begin{equation}
 {\bf H'}={\bf H}-\frac{2}{c}[{\bf v}\times{\bf g}_\mathrm{N}].  
\end{equation}
Since ${\bf H'}=0$ in the comoving reference frame, the gravitomagnetic field can be expressed in terms of the Newtonian field required for the non-relativistic approximation as 
\begin{equation}
{\bf H}=\frac{2}{c}[{\bf v}\times{\bf g}_\mathrm{N}].    
\end{equation}
Thus, we derive from (\ref{4b})
\begin{eqnarray}\label{4b2}
\partial_t (g_i g_k) = && \frac{1}{2}[\nabla \times [{\bf v}\times {\bf g}_\mathrm{N}]]_i g_k \nonumber\\&& + \frac{1}{2}[\nabla \times [{\bf v}\times {\bf g}_\mathrm{N}]]_kg_i \nonumber\\&&+ 4\pi G \rho \left({v_i g_k + v_k g_i}\right).
\end{eqnarray}

In an analogous way, the equation for the second moment of the magnetic induction was obtained in the magnetohydrodynamic approximation in \cite{s10}, and the equation for the second moment of the electric field strength was found in the electrohydrodynamic approximation in \cite{s13}.

The isotropy and the negligible thermal fluctuations require the following equilibrium average values of the first and second correlation moments of the field

\begin{equation}\label{6b}
\left\langle{g_l}\right\rangle_0=0, \quad \left\langle{g_{\mathrm{N}l}
g_{\mathrm{N}l}}\right\rangle_0 = \left\langle{g^2}\right\rangle_0
\frac{1}{3}\delta_{lm} = \mathrm{const}.
\end{equation}

In our approximation, the last (non-linear) term in (\ref{4b2}) can be neglected because it has the second order of smallness with respect to the amplitude. Substituting (\ref{6b}) into (\ref{4b2}) we obtain the following linearized equation for the second correlation moment of the gravitational field strength
\begin{equation}\label{7}
\partial_t \left\langle{g_i g_k}\right\rangle = \frac{1}{6}\left({\partial_k v_i + \partial_i v_k - 2\partial_l v_l \delta_{ik} }\right)\left\langle{g^2}
\right\rangle_0.
\end{equation}
By linearizing the Euler equation (\ref{6g}), we obtain
\begin{eqnarray}\label{81}
&& \partial_t v_i + \frac{1}{\rho_0} \partial_k \left( \rho_0 \langle{v_i v_k}\rangle + \frac{1}{3}\delta_{ik} \left\langle{v^2}\right\rangle_0\rho \right. \nonumber\\  &&+ \left. \frac{{\left\langle{g_i
g_k}\right\rangle - \frac{1}{2}\left\langle{g_l g_l}\right\rangle\delta
_{ik}} }{4\pi G}\right) = 0.
\end{eqnarray}
Taking into account (\ref{0}), (\ref{rein}), and (\ref{7}), we differentiate (\ref{81}) with respect to time and obtain an equation for small acoustic oscillations in an isotropic cloud of stars:
\begin{eqnarray}\label{9}\nonumber
\partial_t^2 v_i - \partial_k\partial_l(v_i\delta_{kl}+v_k\delta_{il})\left(\frac{1}{3}\left\langle{v^2}\right\rangle_0 \right. \nonumber\\ \left.-\frac{\left\langle{g^2}\right\rangle_0}{24\pi G\rho_0}\right)-\frac{1}{3}\left\langle{v^2}\right\rangle_0\partial_i\partial_k v_k = 0.
\end{eqnarray}

\section{Cloud of stars stability}
\label{sec:4}

When solving equation (\ref{9}), it is convenient to switch to Fourier components according to the following rule:
\begin{equation}\label{19}
{\bf v}\left({{\bf x},t}\right) =\frac{1}{(2\pi)^4} \smallint d^3 k\, d\omega\,{\bf
v}\left({{\bf k},\omega}\right)e^{i{\bf kx} - i\omega t}.
\end{equation}

Substituting (\ref{19}) into (\ref{9}), we obtain dispersion equations for longitudinal acoustic oscillations:
\begin{equation}\label{10}
\omega^2_{\|}=\left(\left\langle{v^2}\right\rangle_0 -
\frac{\left\langle{g^2} \right\rangle_0}{12\pi G\rho_0}\right)k^2,
\end{equation}
and for two branches of transverse acoustic oscillations:
\begin{equation}\label{101}
\omega^2_{\bot}=\frac{1}{3}\left(\left\langle{v^2}\right\rangle_0-\frac{\left\langle{g^2}
\right\rangle_0}{8\pi G\rho_0}\right)k^2.
\end{equation}

Taking into account correlations of the gravitational field and the turbulent-like flows in stable cloud in a stationary state, the equation (\ref{10}) gives the following relation for the modified velocity of the longitudinal acoustic wave:
\begin{equation}\label{11}
u_{\|} = \sqrt {\left\langle{v^2}\right\rangle_0 -
\frac{\left\langle{g^2} \right\rangle_0}{12\pi G\rho_0}},
\end{equation}
and equation (\ref{101}) gives the velocity of the transverse acoustic wave:
\begin{equation}\label{112} u_{\bot} = \sqrt{\frac{1}{3}\left(\left\langle{v^2}\right\rangle_0 -\frac{\left\langle{g^2}\right\rangle_0}{8\pi G\rho _0}\right)}.
\end{equation}
If the frequency of perturbation $\omega$ in (\ref{10}) or in (\ref{101}) becomes imaginary, the corresponding mode will increase. 

Expressing (\ref{11}) through (\ref{112}) as
\begin{equation}\label{u-u}
u_{\|} = \sqrt {\frac{1}{3}\left\langle{v^2}\right\rangle_0 +2 u^2_{\bot}},
\end{equation}
one can observe that $u_{\|}$ always exists when $u_{\bot}$ exists. Therefore, it is the transverse perturbations (\ref{101}) that play a crucial role in determining the stability of the system. The stability condition thus follows from (\ref{112})
\begin{equation}\label{102}
\left\langle{v^2}\right\rangle_0 - \frac{\left\langle{g^2}\right\rangle_0}{8\pi G\rho_0} \geq 0.
\end{equation}

The energy density of the gravitational field is given by \cite{ll2}
\begin{equation}
W=-\frac{g^2}{8\pi G}.
\end{equation}
Hence, the cloud is stable when the following condition is satisfied
\begin{equation}\label{103}
\left\langle{v^2}\right\rangle_0 \rho_0 = 2T \geq -W,
\end{equation}
where $T=\frac{1}{2}\left\langle{v^2} \right\rangle_0 \rho_0$ represents the average kinetic energy density.

Thus, the stability limit of the cloud of stars, where $u_{\bot}=0$, coincides with the virial theorem $2T=-W$ (see e.g., \cite{b-s}, Eq. (4--81), \cite{king}, Eq. (9.11)). In this limit, transverse waves do not exist, and the speed of longitudinal sound is given by $u_{\|}=\sqrt{\frac{1}{3}\left\langle{v^2}\right\rangle_0}$.

If the condition (\ref{102}) is not satisfied, transverse modes may increase the kinetic energy until the equality $u_{\bot}=0$ is reached.

By using (\ref{102}), it is also possible to estimate the size of the stable cloud of stars. For the sake of simplicity, let us assume that the cloud has a spherical shape. The gravitational energy of the spherical cloud with the radius $R$ and the mass $M=\frac{4}{3}\pi\rho_0 R^3$ is given by
\begin{equation}
U=-\int_0^R \frac{G}{r} MdM=-\frac{16}{15} G\rho_0^2\pi^2R^5.
\end{equation}
Its density can be estimated as
\begin{equation}
W=-\frac{4}{5}G\rho_0^2\pi R^2.
\end{equation}
Then, the condition (\ref{102}) imposes the following restriction on the radius of the gravitationally bound spherical cloud:
\begin{equation}
\left\langle{v^2}\right\rangle_0 + \frac{W}{\rho_0}=\left\langle{v^2}\right\rangle_0 - \frac{4}{5}G\rho_0\pi R^2 \geq 0,
\end{equation}
from which one can derive:
\begin{equation}\label{104}
R\leq\frac{\sqrt{5\left\langle{v^2}\right\rangle_0}}{2\sqrt{\pi G
\rho_0}}.
\end{equation}
This condition is analogous to the Chandrasekhar condition (\ref{cha}), incorporating turbulence (see \cite{cha}, Eq. (24)).

Taking $\lambda_C$ (\ref{cha}) as the limiting diameter of the stable spherical cloud of stars, we obtain the Chandrasekhar radius \cite{cha}:
\begin{equation}\label{104C}
R_C=\frac{\lambda_C}{2} \leq
\frac{\sqrt{\pi\left\langle{v^2}\right\rangle_0}}{2\sqrt{3 G
\rho_0}}.
\end{equation}
The corresponding expression for the energy of the gravitational interaction is given by:
\begin{equation}\label{WC}
W_C=G\rho_0^2\pi \frac{{\pi\left\langle{v^2}\right\rangle_0}}{{15
G\rho_0}}=\rho_0\frac{{\pi^2\left\langle{v^2}\right\rangle_0}}{{15}}.
\end{equation}
However, this energy does not satisfy the virial theorem:
\begin{equation}\label{VT}
2T=\left\langle{v^2}\right\rangle_0\rho_0 \neq
W_C=\rho_0\left\langle {v^2}\right\rangle_0\frac{\pi^2}{{15}}.
\end{equation}

The relation (\ref{104}) allows us to adjust the Jeans mass $M_J$ based on the Chandrasekhar condition (\ref{104C}) (see \cite{b-s}, Eq. (5--24)):
\begin{equation}\label{MJ}
\frac{M_J}{M_C}=\frac{R^3}{R^3_C}=\left(\frac{\sqrt{15}}{\pi}\right)^{3}\approx 1.87.
\end{equation}
This adjustment increases the mass of the stable gravitating system and is much greater than the weak relativistic corrections found in \cite{k21}.

To evaluate the agreement between the obtained results and observational data, we analyze data from distant massive globular clusters as well as selected old open clusters.

Table~\ref{table1} presents compiled data from the Messier catalog \cite{freecharts}, SIMBAD astronomical database \cite{SIMBAD}, and HyperLEDA database \cite{HyperLEDA} for globular clusters. This data is utilized to estimate the mean square velocities of stars, denoted as $\sqrt{\left\langle{v^2}\right\rangle}_\mathrm{est}$, based on the condition (\ref{104}). The observed values, $\sqrt{\left\langle{v^2}\right\rangle}_\mathrm{obs}$, are extracted from \cite{Illingworth,Gunn,Peterson,Harris} and the Holger Baumgardt globular cluster database \cite{Baumgardt}. 

\begin{table*}[t]
\centering
\caption{\label{table1} 
Structure parameters of selected massive distant globular clusters. M is the cluster mass, $\tau$ is the estimated cluster age, R is the average cluster radius, N is the estimated number of stars in the cluster, $\sqrt{\left\langle{v^2}\right\rangle}_\mathrm{obs}$ represents the observed mean square velocity of stars in the cluster, and $\sqrt{\left\langle{v^2}\right\rangle}_\mathrm{est}$ represents the estimated mean square velocity of stars calculated using (\ref{104})}
\begin{tabular*}{\textwidth}{@{\extracolsep{\fill} }lcccccc}
\toprule
Name & M $[10^5\mathrm{M}_\odot]$ & $\tau$ $[\mathrm{Gyr}]$ & R $[\mathrm{pc}]$  & N $[10^5]$ & $\sqrt{\left\langle{v^2}\right\rangle}_\mathrm{obs}$ $[\mathrm{\frac{km}{s}}]$  & $\sqrt{\left\langle{v^2}\right\rangle}_\mathrm{est}$ $[\mathrm{\frac{km}{s}}]$
\vspace{0.7 mm}\\
\toprule
NGC 6121 & 8.71 & 12.2 & 11.55 & 0.3 & 2.0  & 2.6 \\
NGC 6864 & 3.70 & >13  & 20.54 & 0.4 & 5.0 & 7.1  \\
NGC 6218 & 1.10 & 13.8 & 11.34 & 0.7 & 5.0  & 4.0 \\
NGC 6779 & 2.30 & 13.7 & 12.88 & 0.8 & 4.5 & 4.0 \\
NGC 7078 & 6.33 & 12.5 & 26.98 & 1.0 & 9.0 & 3.0 \\
NGC 7089 & 6.20 & >13  & 26.67 & 1.5 & 8.2  & 3.8 \\
NGC 6266 & 12.2 & 11.8 & 15.02 & 1.5 & 13.7 & 5.1 \\
NGC 1851 & 3.18 & 9.20 & 19.93 & 2.0 & 7.9  & 5.1 \\
NGC 6093 & 3.38 & 12.5 & 14.72 & 2.0 & 9.0 & 5.9 \\
NGC 7006 & 3.03 & >10  & 21.55 & 2.5 & 3.0  & 5.5 \\
NGC 5272 & 4.10 & 11.4 & 27.29 & 5.0 & 4.8  & 6.9 \\
NGC 5024 & 8.26 & 12.7 & 33.74 & 5.0 & 3.8  & 6.2 \\
NGC 0104 & 7.00 & 13.6 & 18.40 & 5.0 & 10.5 & 8.4 \\
NGC 5904 & 8.57 & >13  & 24.53 & 5.0 & 5.7  & 7.3 \\
NGC 2808 & 14.2 & 12.5 & 19.62 & 10.0 & 14.2 & 11.5\\
NGC 6715 & 17.8 & >13  & 46.91 & 10.0 & 14.0 & 7.4 \\
NGC 6254 & 1.89  & 11.4  & 12.26 & 1.0 & 6.6  & 4.6 \\
NGC 6205 & 6.00 & 11.65 & 25.75 & 3.0 & 7.1 & 5.5 \\
NGC 6402 & 6.00 & 13.0  & 14.72 & 1.5 & 11.1 & 5.1\\
\bottomrule
\end{tabular*}
\end{table*}

\begin{table*}[t]
\centering
\caption{\label{table2}
 Structure parameters of selected old open clusters. M, $\tau$, R, and N represent the cluster mass, estimated age, average radius, and estimated number of stars, respectively. $\sqrt{\left\langle{v^2}\right\rangle}_\mathrm{disp}$ and $\sqrt{\left\langle{v^2}\right\rangle}_\mathrm{est}$ denote the observed values from \cite{Pang,Chumak} and the estimated values obtained using (\ref{104}) for the mean square velocities of stars in the cluster}
\begin{tabular*}{\textwidth}{@{\extracolsep{\fill} }lcccccc}
\toprule
Name & M $[\mathrm{M}_\odot]$ & $\tau$ $[\mathrm{Gyr}]$ & R $[\mathrm{pc}]$  & N & $\sqrt{\left\langle{v^2}\right\rangle}_\mathrm{disp}$ $[\mathrm{\frac{km}{s}}]$  & $\sqrt{\left\langle{v^2}\right\rangle}_\mathrm{est}$ $[\mathrm{\frac{km}{s}}]$
\vspace{0.7 mm}\\
\toprule
NGC 1245 & 434 & 1.06 & 3.26 & 358 & 0.70  & 0.54 \\
NGC 1798 & 796 & 1.45 & 3.14  & 580 & 1.02 &  0.70 \\
NGC 2158 & 242 & 2.20 & 3.84  & 200 & 1.82  & 0.37\\
NGC 2420 & 524 & 2.30 & 3.78  & 451 & 0.49 & 0.56 \\
NGC 2682 & 820 & 5.30 & 3.20 & 933 & 0.78 & 0.87 \\
NGC 6791 & 1214 & 9.00 & 6.65 & 1278 & 1.29  & 0.71 \\
NGC 7789 & 915 & 1.70 & 3.96  & 791 & 0.85 &  0.72 \\
Berkeley 17 & 393 & 9.00 & 4.65 & 413 & 0.45  & 0.48 \\
Trumpler 5 & 4313 & 4.95 & 3.25  & 3755 & 1.66 & 1.73 \\
NGC 0188 & 380 & 7.00 & 3.62 & 333 & 0.50  & 0.49 \\
\bottomrule 
\end{tabular*}
\end{table*}

\begin{figure}[t]
\centering
    \begin{minipage}{\linewidth}
        \includegraphics[width=\linewidth]{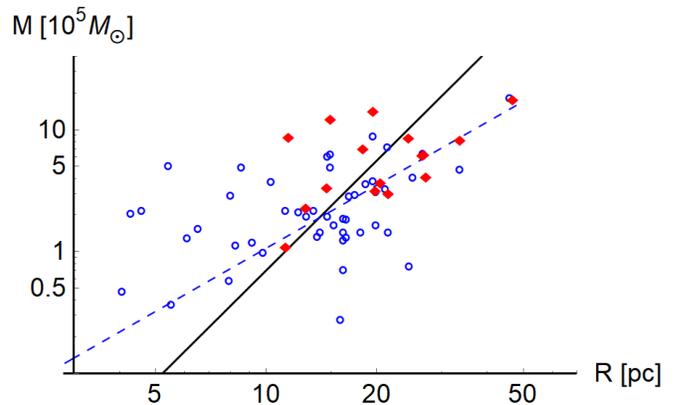}
    \end{minipage}
\caption{Cluster mass as a function of cluster radius. Circles represent data from HyperLeda and Harris \cite{Harris} catalogues. Diamonds indicate clusters listed in Table~\ref{table1}. The dashed line represents the best-fit empirical law $M=0.02R^{1.725}$. The solid line, plotted for diamonds only, corresponds to $M=\frac{4}{3}\pi \left\langle{\rho}\right\rangle R^3$}
\label{fig1}
\end{figure}

Figure~\ref{fig1} shows observed cluster masses and cluster radii from data compilations of HyperLeda \cite{HyperLEDA}, Harris \cite{Harris} catalogues (circles) and Messier catalogue \cite{freecharts}, SIMBAD astronomical database \cite{SIMBAD} (diamonds). The combined data set yields the best fit empirical law $M=0.02R^{1.725}$ indicated by the dashed line. For distant clusters (represented by diamonds according to the data presented in Table~\ref{table1}), the mass-radius relation is mainly governed by the virial theorem. The best fit plotted for the massive distant clusters only (the solid line) follows the usual relation $M=\frac{4}{3}\pi \left\langle{\rho}\right\rangle R^3$. The average cluster density, determined from this relation, is found to be $\left\langle{\rho}\right\rangle=16.50 M_\odot \mathrm{pc^{-3}}$.

It should be noted that the correspondence between the estimated and observed values of mean square velocities of stars in globular clusters, along with realistic estimations of the average cluster density, provides evidence in favor of the consistency of the obtained hydrodynamic model in describing massive distant globular clusters.

\begin{figure}[t]
\centering
    \begin{minipage}{\linewidth}
        \includegraphics[width=\linewidth]{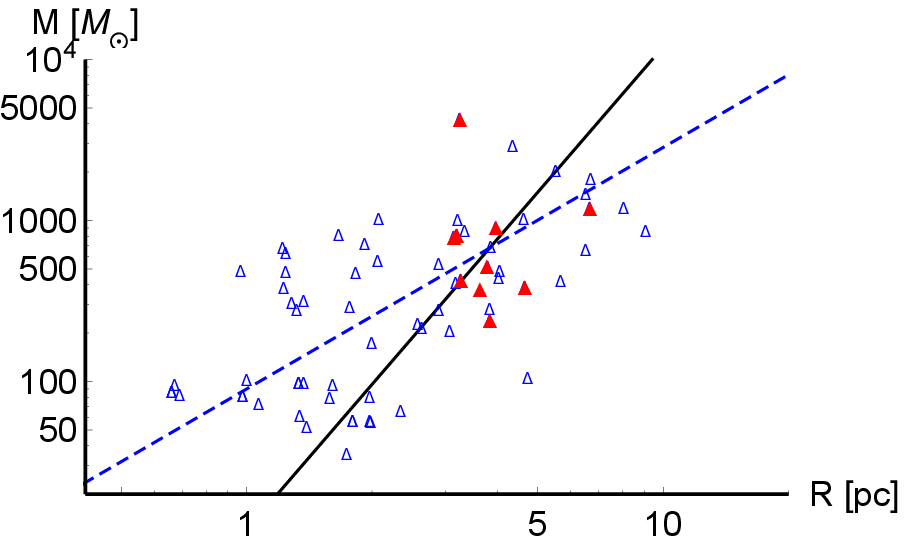}
    \end{minipage}
\caption{Cluster mass as a function of cluster radius. Empty triangles represent data from \cite{Tad}. Solid triangles indicate open clusters listed in Table~\ref{table2}. The dashed line corresponds to the best-fit empirical law $M=90.06R^{1.501}$. The solid line is plotted for solid triangles only following the usual relation $M=\frac{4}{3}\pi \left\langle{\rho}\right\rangle R^3$}
\label{fig2}
\end{figure}

In contrast to globular clusters, open clusters exhibit irregular shapes and contain populations ranging from several hundreds to thousands of member stars. These stars are formed from the same giant molecular cloud, resulting in a roughly homogeneous and isotropic distribution. We use data from \cite{Tad} for older and well-populated open clusters. In our analysis, we approximate open clusters as spheres following \cite{Pang3D}, and apply the condition (\ref{104}) to estimate the mean square velocities of member stars for each selected cluster. The results are presented in Table~\ref{table2}. For comparison, we primarily utilize velocities $\sqrt{\left\langle{v^2}\right\rangle}_\mathrm{disp}$ estimated using line-of-sight velocity dispersion from \cite{Pang}, along with data from \cite{Chumak} for NGC 0188. It is evident that our approach yields excellent agreement with known results for open clusters.

Figure~\ref{fig2} displays observed masses and radii of open clusters. The data obtained from \cite{Tad} is represented by empty triangles. Solid triangles indicate older clusters listed in Table~\ref{table2}. The dashed line represents the best-fit empirical mass-radius relation for all depicted open clusters, given by $M=90.06R^{1.501}$. The best fit plotted for the older clusters from Table~\ref{table2}  only (the solid line) follows the usual relation $M=\frac{4}{3}\pi \left\langle{\rho}\right\rangle R^3$. This relation yields an average open cluster density of $\left\langle{\rho}\right\rangle=2.94 M_\odot \mathrm{pc^{-3}}$, which is consistent with current observations.

\section{Conclusions}
\label{conc}
The presence of analogs of anisotropic pressure, along with the turbulent-like motion of stars within a gravitating homogeneous and isotropic star cloud, gives rise to transverse perturbations. Accounting for these transverse perturbations when studying the stability of a gravitating system leads to a more appropriate generalization of the Jeans relation.

In this paper, we develop a hydrodynamic model that describes small oscillations in a cloud of stars based on the Jeans equations. By considering the non-relativistic limit of the 1PN approximation of the Einstein equations and including the transverse component of the gravitational field strength, we derive expressions for one longitudinal and two transverse branches of acoustic oscillations within the cloud.

It is revealed that transverse oscillations disrupt the stability of the system at shorter wavelengths compared to Chandrasekhar's estimation (\ref{cha}) which does not take into account the transverse component of the gravitational field strength. When the velocity of transverse oscillations equals zero, a stability condition arises for the cloud of stars, which requires the kinetic energy density to be equal to half of the gravitational interaction energy density, precisely corresponding to the virial theorem.
 
Based on our analysis, we conclude that the estimated radius of a stable spherical cloud of stars must satisfy the generalized relation (\ref{104}). Applying this result to observational data for globular and open star clusters demonstrates a strong correspondence with current observations.

\section*{acknowledgement}
The authors acknowledge the Research Centre for Theoretical Physics and Astrophysics, Institute of Physics, Silesian University in Opava for institutional support. I.B. acknowledges the project SGS/26/2022. E.K. acknowledges the Physics Department of the University of Illinois at Urbana-Champaign for the hospitality, institutional and existential support.

\end{document}